\documentclass[prl,aps,twocolumn,superscriptaddress,notitlepage]{revtex4-1}
\usepackage{amssymb,amsfonts,amsmath,amstext,graphicx,hyperref}
\usepackage{tikz}
\usepackage{graphicx}
\usepackage[normalem]{ulem}
\usepackage{mathtools}
\hypersetup{
	colorlinks=true,
	linkcolor=blue,
	filecolor=magenta,      
	urlcolor=cyan,
}
\usepackage{xcolor}

\usepackage[normalem]{ulem}
\newcommand{\cut}[1]{}

\usepackage{physics}

\begin{document}
\title{Amplification, Mitigation and Energy Storage via Constrained Thermalization}
\author{Midhun Krishna}\thanks{equal contribution}
\affiliation{Department of Physics, Indian Institute of Technology-Bombay, Powai, Mumbai 400076, India}

\author{Harshank Shrotriya}\thanks{equal contribution}

\affiliation{Centre for Quantum Technologies, National University of Singapore, 3 Science Drive 2, Singapore 117543}
\author{Leong-Chuan Kwek}
\affiliation{Centre for Quantum Technologies, National University of Singapore, 3 Science Drive 2, Singapore 117543}
\affiliation{MajuLab, CNRS-UNS-NUS-NTU International Joint Research Unit, UMI 3654, Singapore}
\affiliation{National Institute of Education,
Nanyang Technological -  University, 1 Nanyang Walk, Singapore 637616}
\affiliation{School of Electrical and Electronic Engineering
Block S2.1, 50 Nanyang Avenue, 
Singapore 639798 }
\author{Varun Narasimhachar}
\affiliation{Institute of High Performance Computing (IHPC), Agency for Science Technology and Research (A*STAR), 1 Fusionopolis Way, \#16-16 Connexis, Singapore 138632, Republic of Singapore}
\author{Sai Vinjanampathy}
\email{sai@phy.iitb.ac.in}
\affiliation{Department of Physics, Indian Institute of Technology-Bombay, Powai, Mumbai 400076, India}
\affiliation{Centre of Excellence in Quantum Information, Computation, Science and Technology, Indian Institute of Technology Bombay, Powai, Mumbai 400076,India}
\affiliation{Centre for Quantum Technologies, National University of Singapore, 3 Science Drive 2, Singapore 117543}

\begin{abstract}
Amplification (mitigation) is the increase (decrease) in the change of thermodynamic quantities when an initial thermal state is thermalized to a different temperature in the presence of constraints, studied thus far only for permutationally invariant baths. In this manuscript, we generalize amplification and mitigation to accommodate generic strong symmetries of open quantum systems and connect the phenomenon to Landauer's erasure. We exemplify our general theory with a new bath-induced battery charging protocol that overcomes the passivity of KMS-preserving transitions.
\end{abstract}

\maketitle

\paragraph{Introduction.---}
Thermalization is a fundamental process describing how physical systems, from classical gases to quantum many-body systems, tend toward equilibrium.
Various studies have investigated thermalization in both closed \cite{goldstein_canonical_2006,rigol2008thermalization} and open \cite{vznidarivc2010thermalization,reichental2018thermalization} quantum systems.
The thermalization of subsystems of closed systems has been studied from various approaches such as typicality \cite{goldstein_canonical_2006,rigol2008thermalization}.
Closed systems are often described by Hamiltonians that exhibit discrete or continuous symmetries \cite{ilievski2016quasilocal,vidmar_generalized_2016} and these approaches to study thermalization have been extended to such systems with conserved quantities \cite{jaynes1957information,jaynes1957information2,popescu2006entanglement, linden2009quantum,polkovnikov2011colloquium,ilievski2016quasilocal,vidmar_generalized_2016,yunger2016microcanonical,lenard1978thermodynamical,bera2019thermodynamics,mitsuhashi2022characterizing,khanian2023resource}.
The thermalization in the presence of conserved quantities prevents complete equilibriation of quantum systems leading to generalized Gibbs ensembles, even with noncommuting charges \cite{yunger2016microcanonical,murthy_non-abelian_2022,kranzl2023experimental}.

The alternate approach to study thermalization is the dynamical equilibration of open systems described using microscopically motivated completely positive and trace-preserving (CPTP) maps or master equations to model explicitly thermal as well as non-thermal baths \cite{breuer2002theory,lidar2019lecture}. 
Like in the case of closed quantum systems, symmetries in open quantum systems
are well studied in various physically relevant models \cite{buca12,albert2014symmetries,vznidarivc2013coexistence,van2018symmetry,manzano2018harnessing,manzano2014symmetry,manzano2018harnessing,manzano2021coupled,thingna2021degenerated,cattaneo2020symmetry,halati2022breaking,groot,Aash,lau2023convex,lieu2020symmetry,groot,verissimo2023dissipative,Buca_non_abelian,roberts2021hidden,prosen2012p,altland2021symmetry,kawabata2023symmetry} and they have found diverse applications \cite{Aash,lau2023convex,manzano2014symmetry} such as passive error correction \cite{lieu2020symmetry} and in preserving topological order \cite{groot,verissimo2023dissipative}.
However, the studies of open quantum systems thermalization in the presence of symmetries are limited, in particular to the case of indistinguishable noninteracting particles driven by collective interaction with the bath \cite{latunePRA, LaturePRR,kamimura_quantum-enhanced_2022,Ben,jaseem2023collective,jaseem2023quadratic}. 
Such collective models have been well studied in the literature as they host various exciting physics \cite{garraway2011dicke}, including Dicke superradiance \cite{dicke1954coherence} and time crystals \cite{iemini2018boundary}, and applications \cite{vetrivelan2022near,dou2022charging}.
Collective thermalization has been shown to exhibit amplification or mitigation effects of thermodynamic quantities when compared with equilibrium values, which informs the design of quantum thermal machines with better performance. 
These deviations for various thermodynamic quantities have been characterized \cite{LaturePRR}, and applied to quantum thermometry \cite{latune_collective_2020} and enhance work output \cite{latune_collective_2020,LaturePRR,kamimura_quantum-enhanced_2022,Ben,jaseem2023collective}, and improve reliability \cite{jaseem2023quadratic} of thermal machines.

The permutational invariance of collective models where thermalization has been studied before is an example of a so-called strong symmetry, discussed below. 
Besides permutational invariance, various other examples of physically well-motivated dynamics with strong symmetry are well known in the literature \cite{buca12,albert2014symmetries,vznidarivc2013coexistence,van2018symmetry,manzano2018harnessing,manzano2014symmetry,manzano2018harnessing,manzano2021coupled,thingna2021degenerated,cattaneo2020symmetry,halati2022breaking,groot} and have vast applications in the physics of quantum matter. 
What is lacking is a systematic study of thermalization and its interplay with information theory in such generic open quantum systems with strong symmetries. 
In this work, we generalize the study of thermalization to go beyond permutational symmetry and discuss the \textit{thermalization of generic strong symmetries}, or ``s-thermalization'' for short. 
We formulate thermalization for generic strong symmetries, 
thus paving the path for designing thermal machines. 
As an outcome, our main result is to show how amplification and mitigation can be understood as an imbalance of erasure entropy providing an information theory perspective on such reservoir engineering.
We illustrate our formalism by showcasing how constrained thermalization can be employed as a mechanism for quantum battery charging.

We begin with a brief review of strong symmetries of open quantum system dynamics and discuss thermalization in the presence of these symmetries and conserved quantities.
Thermalization in the presence of conserved quantities leads to the preservation of the initial state symmetry structure, which we implement by employing a Maxwell demon. 
Such a demon monitors the symmetries of the system and uses the information gained to perform constrained thermalization within the symmetry sectors.  
Identifying Maxwell demon with symmetry admits an interpretation for the deviation of steady-state internal energy from that of generic thermalization in terms of Landauer's erasure. Finally, we exemplify the potential of constrained thermalization by discussing an application in the context of quantum battery charging.

\paragraph{Symmetry-Constrained Thermalization.---} 
Symmetries of open quantum systems are distinct from those of closed quantum systems in many ways, starting from a lack of one-to-one correspondence between continuous symmetries and conserved quantities \cite{buca12,albert2014symmetries}. 
Open systems can have so-called \textit{weak} and \textit{strong} symmetries. An open quantum system is said to have a strong symmetry when the symmetry operators commute with all the Kraus operators (in the case of dynamical maps) \cite{groot} or with the Hamiltonian and Lindblad operators (in the case of continuous Markovian evolution) \cite{buca12,albert2014symmetries,manzano2018harnessing}. Strong symmetries have been studied in various models, including driven spin chains \cite{buca12,vznidarivc2013coexistence,van2018symmetry,manzano2018harnessing}, qubit networks and molecular systems \cite{manzano2014symmetry,manzano2018harnessing,manzano2021coupled,thingna2021degenerated}, bosonic \cite{cattaneo2020symmetry,halati2022breaking} and fermionic \cite{cattaneo2020symmetry} systems, and noisy channels \cite{groot}. 
The Hilbert space of such systems possessing a strong symmetry 
can be decomposed as $\mathcal{H} = \bigoplus_i \mathcal{H}_i$, where each $\mathcal H_i$ is an invariant subspace of the symmetry, corresponding to the eigendecomposition of the symmetry operator \cite{thingna2021degenerated}.
The trace-preserving property of the evolution implies that each of these invariant subspaces (in the Hilbert space) is associated with at least one fixed point ($\rho_i$) of the propagator (in the Liouville space) \cite{buca12,albert2014symmetries,thingna2021degenerated}. 
Thus, the system as a whole has multiple steady states $\{\rho_i\}$ spanning the so-called
steady-state manifold. 
Starting from an arbitrary initial state $\rho$, the steady states obtained from dissipative dynamics can be written as $\rho_\text{SS} = \bigoplus_i p_i \rho_i$.
Here, $p_i = \Tr(\mathbb{P}_i \rho_0)$ are the initial probabilities corresponding to the initial state $\rho_0$ and $\mathbb{P}_i$ denoting the projector onto
the invariant subspace $\mathcal{H}_i$.
The presence of strong symmetries thus 
preserves the memory of the initial state during dynamics, and we focus on steady states of such dynamics driven by a single thermal bath. 

Owing to Evan's theorem, the interaction of a system with a thermalizing bath (satisfying the Kubo\textendash Martin\textendash Schwinger condition \cite{breuer2002theory,lidar2019lecture}) without any strong symmetry leads to generic thermalization  \cite{breuer2002theory,buca2015transport}. 
Generic thermalization of a system described by Hamiltonian $H$ in contact with a bath at inverse temperature $\beta$ results in the steady state of the Gibbs form, $\omega^{\beta} = e^{-\beta H}/\Tr(e^{-\beta H}) $.
In contrast, strong symmetric interaction with such a thermalizing bath leads to thermalization within the symmetry subspaces \cite{yoshida2023uniqueness}, thus inducing a nonequilibrium steady state that preserves a part of the initial state information. 
The steady states have the form $\rho_\text{SS} = \bigoplus_i p_i \omega_i^{\beta}$, with  $p_i$ being the initial state probabilities preserved by the strong symmetry and $\omega_i^{\beta} = e^{-\beta \mathbb{P}_iH\mathbb{P}_i}/\Tr(\mathbb{P}_i e^{-\beta H})$ being the steady states of the dynamics having the Gibbs form, both corresponding to invariant subspace $\mathcal{H}_i$.

\begin{figure*}[htp!]
    \centering
    \includegraphics[width=\linewidth]{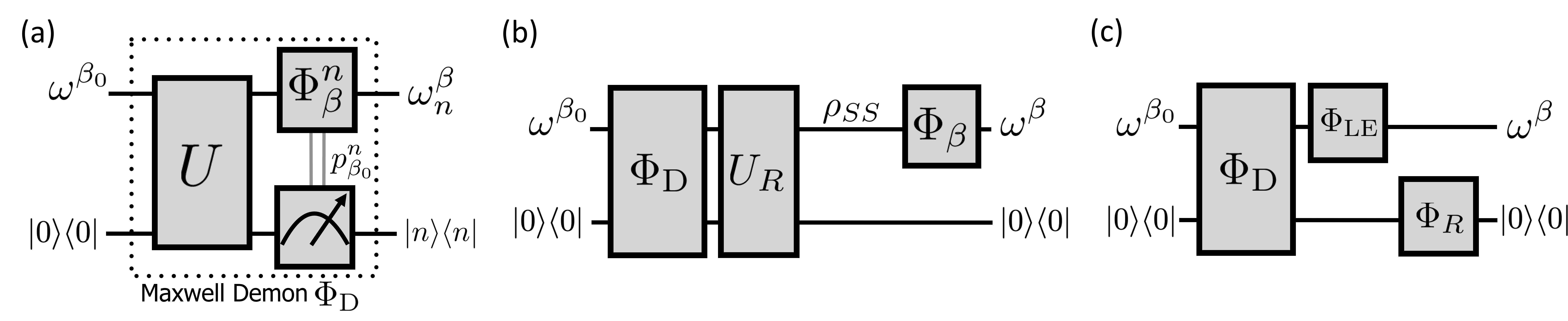}
    \caption{A circuit diagrammatic representation of various thermalization mechanisms discussed in the manuscript. (a) s-thermalization process depicted by the map $\Phi_\text{D}$ in which Maxwell demon correlates symmetry sectors of the input thermal state with a register and assists in constrained thermal thermalization by using the outcome of measurements on the register. (b) and (c) show the two pathways towards generic thermalization. (b) depicts the pathway where the information in the demon's register is erased first by the controlled unitary $U_R$, resetting it to the fiducial state using the symmetry of the system as control. This mixes states from various symmetry sectors to yield $\rho_\text{SS}$, which is further acted upon by the thermal map $\Phi_{\beta}$ leading to a generic thermalization. (c) represents full generic thermalization as a composition of $\Phi_\text{D}$ and Landauer's erasure (depicted as $\Phi_\text{LE}$), which is applied on the output of the s-thermalization to erase information encoded in the thermal ensemble and finally, the map $\Phi_{R}$ resets the register.}
    \label{fig:cct_erasure}
\end{figure*}

\paragraph{S-Thermalization as Maxwell demon.---} 
To elucidate the role of symmetry in s-thermalization, we introduce a Maxwell demon who is an agent that explicitly brings out the effect of symmetry. The demon does this by correlating the state space corresponding to $\mathcal{H}_i$ with state $|i\rangle \langle i|$ of an ancilla register. 
In our scenario, the demon does not perform feedback but instead is a tool to understand the role of the symmetry constraint and the role of the constrained thermalization in the nonequilibrium thermodynamic properties of the system. 
We illustrate the Maxwell demon implementation of s-thermalization by a thermalizing bath at inverse temperature $\beta$ from an initial state $\omega^{\beta_0}$ in Fig.~\ref{fig:cct_erasure}(a). 
Initially, the demon acquires information about the symmetry structure of the state by correlating the symmetry degree of freedom with the register initialized in the reference state $\ket{0}\bra{0}$, using unitary $U = \sum_n \mathbb{P}_n \otimes \left(\ket{n}\bra{0}+\ket{0}\bra{n} + \sum_{k\neq n,0} \ket{k}\bra{k}\right)$, and performing projective measurement of the register.
The step is accompanied by a decrease in the entropy of the system (denoted as $S(\rho) \coloneqq -\text{Tr}(\rho \ln{\rho})$) since $S(\omega^{\beta_0}) \geq \sum_i p_i^{\beta_0}S(\omega_i^{\beta_0})$ and
the difference is exactly the information-theoretic entropy $H(p^{\beta_0}) = -\sum_i p_i^{\beta_0} \log p_i^{\beta_0}$ associated with the information stored in the memory.
The information about the initial probabilities of symmetry subspaces is now with the Maxwell demon ($\{p_i^{\beta_0}\}$), which can be exploited to perform controlled thermalization, $\Phi_\beta^{n}$, conditioned on the measurement outcome. 
This application can be contrasted to a protocol where the information acquired by demon is utilized to increase work extraction via feedback 
\cite{quan_maxwells_2006,jacobs_second_2009,jacobs_quantum_2012,vinjanampathy2016quantum,elouard_extracting_2017,engelhardt_maxwells_2018}. Note that in such a protocol the optimal improvement gained in extracted work is $H(p^{\beta_0})$.
The Maxwell demon generate states $\omega_{i}^{\beta} \otimes  \ket{i}\bra{i}$ with probabilities $p_i^{\beta_0}$, thereby encoding these probabilities in the subspace restricted states $\omega_{i}^{\beta}$.
In order to quantify the excess nonequilibrium energy in the s-thermalized state, we erase the memory of the register by unitary resetting (via $U_{R} = U^{\dagger}$ in Fig~\ref{fig:cct_erasure}(b)).

The reset step can be thought of as transferring the entropy $H(p^{\beta_0})$ back to the system from the memory register, accompanied by a loss of mutual information between the system and the memory register. 
Note that the entropy change during $U_{R}$ is zero since the reset is a controlled unitary with system states as the control and memory states as the target. 
Additionally, assuming that the energy levels of the register are degenerate, the reset $U_{R}$ is not accompanied by any work cost either. 
The reset $U_R$ leaves the system in the s-thermalized state, $\rho_\text{SS}$, which has deviations in thermodynamics quantities from the thermal state, $\omega^{\beta}$, in equilibrium with the bath.
These deviations have been qualified as amplification (mitigation) when the deviation in the quantities is increased (reduced) in steady states of s-thermalization compared to generic thermalization \cite{LaturePRR}. 
Amplification/mitigation of such deviations can be quantified by studying the change of these quantities when the system is let to fully thermalize with the bath ($\Phi_\beta$ in Fig~\ref{fig:cct_erasure}(b)) by lifting the symmetry constraints. 
The non-equilibrium steady state always has excess free energy, $F(\rho_{\text{SS}})-F(\omega^{\beta}) = S(\rho_{\text{SS}}||\omega^{\beta})/\beta$ \cite{santos2019role}, with respect to the bath. Here, $F(\rho) = E(\rho)-S(\rho)/\beta$ is the free energy of $\rho$ with respect to the bath at inverse temperature $\beta$, and $E(\rho) = \text{Tr}(H\rho)$ is the internal energy. 
It follows that the change in internal energy enabled by the heat exchange with the bath can be expressed as (see appendix A) \begin{equation}
\label{eq:internal_energy_from_F}
    E(\rho_\text{SS}) - E(\omega^{\beta}) = \frac{1}{\beta} \left(S(\rho_\text{SS}) - S(\omega^{\beta}) +S(\rho_\text{SS}||\omega^{\beta})\right).
\end{equation}

The difference in internal energy above can be understood using the information erasure of the quantum ensemble generated by the Maxwell demon. In Fig.~\ref{fig:cct_erasure}(c), the map $\Phi_\text{LE}$ erases the encoded information from the quantum ensemble formed by the Maxwell demon. 
This is accomplished by the thermal randomization of the information \cite{plenio1999holevo,plenio2001physics,maruyama2009colloquium} using the bath at $\beta$ (i.e., $\Phi_\text{LE} = \Phi_{\beta}$) resulting in the composite state $\omega^{\beta} \otimes \sum_i p_i^{\beta_0}\ket{i}\bra{i}$.
The information erasure map $\Phi_\text{LE}$ erases the mutual information between the memory register and the quantum state of the system, and the map $\Phi_{R}$ resets the memory to the reference state, leaving the system unaffected.
This formalism is used below to analyze the amplification-mitigation threshold.

 \begin{figure*}[htp!]
    \centering
    \includegraphics[width=0.9\linewidth]{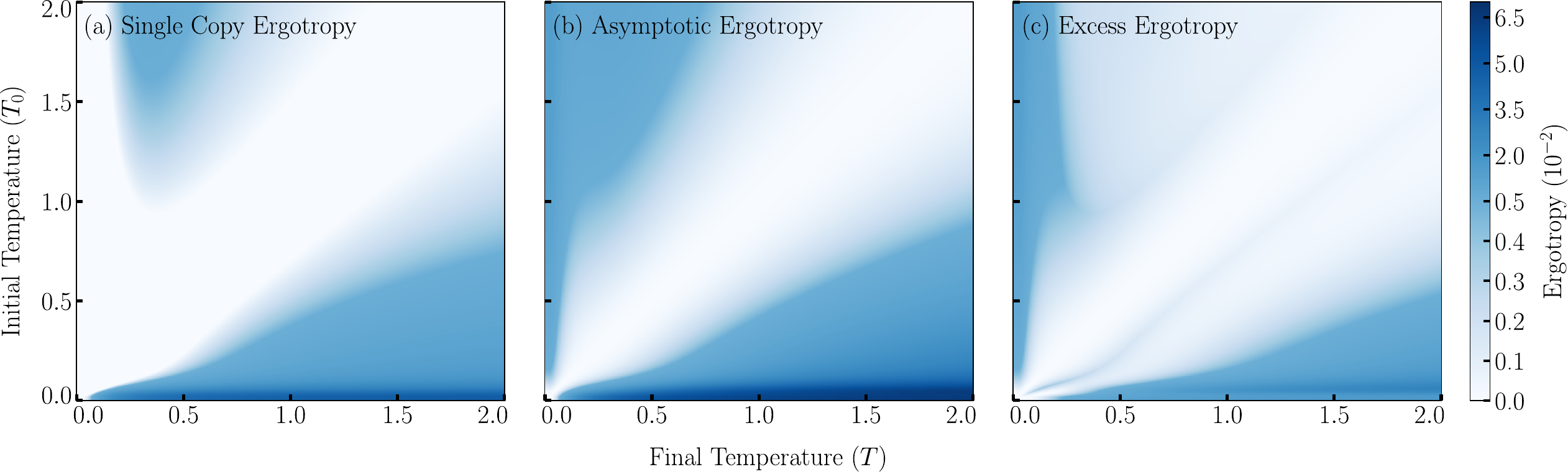}
    \caption{The color plot showing ergotropies of the 4-level system for various initial and final temperatures. The ergotropy of a single copy of the system shows two islands of non-zero value in (a) corresponding to population inversion developed between different levels, as discussed in the main text. The asymptotic ergotropy shows non-zero values for almost all initial and final temperatures except when they are equal, as seen in (b). The excess ergotropy per system of asymptotic number of copies compared to a single copy is shown in (c). The energies of the four levels used to generate the plots are $\{E_0,E_1,E_2,E_3\} = \{0,0.1,0.2,1\}$ and the temperature scales are in the units of $k_b$. The color plot has been plotted on a 2-step linear gradient for better contrast.}
    \label{fig:battery}
\end{figure*}

\paragraph{Example: Mitigation and Amplification as Erasure---} In this section, we study the amplification/mitigation behavior of s-thermalization on internal energy. We focus on internal energy since previous works \cite{LaturePRR,Ben} have related this effect to improvement in the work output of thermal machines. 
The deviation of internal energy given by Eq~(\ref{eq:internal_energy_from_F}) can be interpreted from the information erasure perspective by analyzing the two distinct circuit diagrams (highlighted in Fig.\ref{fig:cct_erasure} (b) and (c)) that follow $\Phi_\text{D}$.
The information encoded in the quantum ensemble after the action of the demon can be erased by a Landauer erasure $\Phi_\text{LE}$ \cite{plenio1999holevo,plenio2001physics,maruyama2009colloquium,goold_role_2016}.
The thermodynamic cost for erasure has the form $\Delta S(\Phi_\text{LE}) = \Delta S_{\text{sys}} + \Delta S_{\text{bath}}$, with $\Delta S_{\text{sys}} = S(\omega^{\beta}) - \sum_i p_i^{\beta_0}S(\omega_i^{\beta})$ being the change in entropy of the system and $\Delta S_{\text{bath}}$ the change in entropy of the bath during $\Phi_\text{LE}$. When heat transfer between the system and bath happens quasi-statically such that no work is done, the cost of erasure can be calculated \cite{plenio1999holevo,plenio2001physics,maruyama2009colloquium,goold_role_2016} as (see appendix B) 
\begin{equation}
\label{eq:erasure_entropy}
    \Delta S(\Phi_\text{LE}) = S(\rho_\text{SS}) - \sum_i p_i^{\beta_0}S(\omega_i^{\beta}) + S(\rho_\text{SS}||\omega^{\beta}).
\end{equation}
The difference of the first two terms above is the extra entropy of erasure ($H(p^{\beta_0})$) that contributes to the increase in free energy when the Maxwell demon possesses information on the population of symmetry sectors. The last term points at the nonequilibrium nature of $\rho_\text{SS}$.
Using the above equation, the difference in internal energy of Eq~(\ref{eq:internal_energy_from_F}) can be rephrased in terms of the cost of erasure as
\begin{equation}
\label{Internal_Energy_from_LE}
    E(\rho_\text{SS}) - E(\omega^{\beta}) = \frac{1}{\beta}\left(\Delta  S(\Phi_\text{LE}) - \Delta S_\text{sys}\right).
\end{equation}
The equation above explicitly expresses the excess internal energy of $\rho_\text{SS}$ in terms of information-theoretic quantities. 
The break-even point of amplification and mitigation can understood as the condition of balanced erasure entropy, $\Delta S_{\text{sys}} = \Delta S(\Phi_\text{LE})$ (see appendix C for a formal definition of the break-even point). 
However, when the above condition is not satisfied, $\Delta S_\text{bath}$ is non-zero indicating an imbalance of erasure entropy ($\Delta S_{\text{sys}} \neq \Delta S(\Phi_\text{LE})$), the sign of which determines if there is amplification or mitigation of steady state internal energy. This information-theoretic understanding of amplification / mitigation is our main result.

Considering the case with $\beta < \beta_0$, where $\beta$ is the inverse bath temperature and $\beta_0$ is the inverse temperature of the initial state, we note that both s-thermalization and general thermalization take the initial state $\omega^{\beta_0}$ to a steady state, $\rho_\text{SS}$ and $\omega^{\beta}$ respectively, with higher internal energy. In this scenario, amplification is said to occur when $E(\rho_\text{SS}) - E(\omega^{\beta}) > 0 \implies \Delta S_\text{bath}>0$ i.e., heat flows from system to bath when the symmetry constraints on the system-bath interaction are removed. On the contrary, mitigation happens when the inequality signs are reversed implying heat flow from bath to system on removing the symmetry constraints. The case with $\beta > \beta_0$ proceeds similarly but with the two kinds of steady states having lower internal energy than the initial state. 

\paragraph{Application: Battery Charging via S-Thermalization.--} We now apply s-thermalization to the question of battery charging \cite{campaioli_quantum_2018,rodriguez_optimal_2022,quach_superabsorption_2022,campaioli_colloquium_2023}. Quantum resources have shown an advantage in the deposited power \cite{binder_quantacell_2015,campaioli_enhancing_2017,campaioli_quantum_2018,rossini_quantum_2020,gyhm_beneficial_2023} and extractable work \cite{alicki2013entanglement,campaioli_quantum_2018} of quantum batteries. 
In this manuscript, we are interested in the extractable work.
Ergotropy, $\mathcal{W} = E(\rho) - \min_{U} E(U\rho U^{\dagger}),$ is defined as the maximum work extractable from a quantum system $\rho$ via unitary transformations $U$, when energy is measured in the eigenbasis of $H$. 
States of the battery with zero and non-zero ergotropy are called passive and active respectively. 
Even though passive states have zero ergotropy, an ensemble of such passive states may allow work extraction. The asymptotic ergotropy \cite{hovhannisyan2013entanglement} defined as the regularized ergotropy per system can be shown to be $S(\rho_\text{SS}||\omega^{\beta^*})/\beta^*$. Here, $\omega^{\beta^*}$ is a Gibbs state having the same entropy as the battery state, i.e., $S(\rho_\text{SS}) = S(\omega^{\beta^*})$.
Thermal states form a subset of passive states called completely passive states for which asymptotic ergotropy is zero.
Hence, thermal resources and baths are typically understood as undesirable for battery charging.   
Here, we propose a battery as the nonequilibrium steady state achieved by s-thermalization. The steady state achieved after s-thermalization does not offer an advantage based on energy coherence or entanglement; however, its utility as a charged battery state can still be demonstrated by virtue of it not being completely passive.

In Fig.~\ref{fig:battery}, we demonstrate this application using as a working model a 4-level system with energy levels labeled by $\{\ket{E_i}\}$ with $0\leq i \leq 3$. Additionally, we assume the presence of a strong symmetry that decomposes the Hilbert space into a direct sum of even and odd parity sectors (an example could be Parity symmetry for systems with two-photon dissipators \cite{albert2014symmetries}). Thus, $\{\ket{E_0}, \ket{E_2}\}$ form the even parity sector and $\{\ket{E_1}, \ket{E_3}\}$ form the odd parity sector and we have chosen energy levels to be such that $E_3-E_1 \gg E_2 - E_0$. The system starts in an initial thermal state at temperature $T_0$ and is charged via s-thermalization using a thermal bath at temperature $T$. As shown, regions of positive ergotropy can be obtained for a wide range of initial temperatures $T_0$ and bath temperatures $T$.

It can be noted that regions of positive ergotropy are enabled by population inversions in steady states post s-thermalization even for the single copy regime, as shown in Fig~\ref{fig:battery}(a). The plot shows two disconnected regions of non-vanishing ergotropy, which can be qualitatively understood as follows.
For $T_0 \ll T$, the s-thermalized state is an active battery when there is a population inversion between energy levels $E_1$ and $E_2$.
Similarly, for $T_0 \gg T$, since $E_3-E_1 \gg E_2 - E_0$, s-thermalization generates a population inversion between energy levels $E_0$ and $E_1$ when populations of $E_0$ and $E_2$ are nearly equal and population of $E_1$ is greater than that of $E_3$.
Further, steady states post s-thermalization are typically not thermal and thus have non-zero asymptotic ergotropy. In the particular case of the 4-level system studied here, the asymptotic ergotropy vanishes only when $T_0 = T$ as seen in Fig~\ref{fig:battery}(b).

\paragraph{Discussion.---} Strong symmetries of open quantum systems decompose a system's Hilbert space into invariant subspaces, modifying the nature of steady states. A specific example of such constrained steady states, namely permutationally invariant steady states, was previously shown to have an advantage in the design of thermal machines \cite{latune_collective_2020,LaturePRR,kamimura_quantum-enhanced_2022,Ben,jaseem2023collective,jaseem2023quadratic}. In this work we present the general theory of amplification and mitigation of thermodynamic quantities under open system constrained by generic strong symmetries. Though we only focused on internal energy as a thermodynamic quantity of interest, other thermodynamic quantities follow from the discussion in a straightforward fashion. Our theory now allows for the design of novel quantum batteries and s-thermal machines in such systems.

Physical systems that conserve total magnetization \cite{buca12,van2018symmetry,manzano2018harnessing,lau2023convex} and parity \cite{albert2014symmetries,lieu2020symmetry} are well known examples of strong symmetries that highlight the general appeal and applicability of our results. Furthermore, we propose design primitives wherein symmetry-driven reservoir engineering of open quantum systems can provide an autonomous design for population inversion, which has several well-known applications including masers \cite{niedenzu2019concepts,rodriguez2023ai}. We contribute to the study of such nanoscale thermal machines by elucidating the thermodynamic meaning of strong symmetries in the context of work extraction/deposition. We hope that this will add to the literature on symmetry driven reservoir engineering which has diverse applications such as battery charging, cooling and entanglement generation.

\begin{acknowledgments}
M.K. acknowledges support by Prime Minister's Research Fellowship (PMRF) offered by the Ministry of Education, Govt. of India. HS acknowledges funding from the Centre for Quantum Technologies (CQT) PhD program. SV acknowledges support from a DST-QUEST grant number DST/ICPS/QuST/Theme-4/2019. 
\end{acknowledgments}

\newpage
\appendix
\maketitle
\onecolumngrid

\section{Appendix A: Internal Energy Difference from Quantum Information Thermodynamics}
In this section, we derive Eq.~(\ref{eq:internal_energy_from_F}) of the main text for the difference in energy between $\rho_\text{SS}$ and $\omega^{\beta}$ and quantify mitigation and amplification of the internal energy in terms of information-theoretic quantities such as entropy and relative entropy.
The non-equilibrium free energy of a quantum state $\rho_\text{SS}$ with respect to a reference bath with inverse temperature $\beta$ can be defined as \cite{santos2019role},
\begin{equation}\label{eq:neqF} 
F(\rho_\text{SS}) = E(\rho_\text{SS}) -  \frac{S(\rho_\text{SS})}{\beta},
\end{equation}
where $S(\rho) = \Tr{\rho \log{ \rho}}$ is the von Neumann entropy of the state $\rho$.
The above equation in terms of the free energy of the equilibrium state at temperature $T$ can be expressed as,
\begin{equation} \label{eq:neqF2} 
F(\rho_\text{SS}) = F(\omega^{\beta}) + \frac{ S(\rho_\text{SS} ||\omega^{\beta})}{\beta} .
\end{equation}
where $S(\rho||\sigma) = -S(\rho) - \Tr{\rho \log{ \sigma}}$ is the quantum relative entropy of $\rho$ with respect to $\sigma$. The free energy function for the equilibrium state has the form,
$ F(\omega^{\beta}) = E(\omega^{\beta}) + S(\omega^{\beta})/\beta = - (\log{Z_{\beta}})/\beta ,$
where $E(\omega^{\beta}) = \Tr{H\omega^{\beta}}$ and $Z_{\beta} = \Tr(e^{-\beta H})$.

We are interested in comparing the steady state energies of the two kinds of thermalization pathways, namely s-thermalization and generic thermalization. Using Eq.~(\ref{eq:neqF}) and Eq.~(\ref{eq:neqF2}), the difference in internal energy of the steady state after evolution by the two pathways becomes,
\begin{equation}
    E(\rho_\text{SS}) - E(\omega^{\beta}) =  \frac{1}{\beta} \left(S(\rho_\text{SS}) - S(\omega^{\beta}) +S(\rho_\text{SS}||\omega^{\beta})\right).
\end{equation}
Thus, we have arrived at the expression for the difference in energy between $\rho_\text{SS}$ and $\omega^{\beta}$ using the definition of free energy of these states with respect to the final bath at inverse temperature $\beta$.

\section{Appendix B: Calculating the Cost of Landauer Erasure}
In this section, we calculate the entropy of erasure for the process $\Phi_\text{LE}$ given by Eq~(\ref{eq:erasure_entropy}) in the main text. The entropy of erasure associated with the process \cite{plenio1999holevo,plenio2001physics,maruyama2009colloquium,goold_role_2016}, $\Delta S(\Phi_\text{LE})$, is a positive quantity given by 
\begin{equation}
\label{eq:erasure_as_sum}
    \Delta S(\Phi_\text{LE}) = \Delta S_{\text{sys}} + \Delta S_{\text{bath}}.
\end{equation} 
Where $S_{\text{sys}}$ and $ \Delta S_{\text{bath}}$ are entropy changes of the system and bath, respectively.
Before the erasure, the system state is an ensemble of symmetry-restricted thermal states $\omega_{i}^{\beta}$ which occurs with probability $p_i^{\beta_0}$ and entropy of this ensemble is $\sum_i p_i^{\beta_0}S(\omega_i^{\beta})$. After erasure the quantum state of the system is $\omega^{\beta}$ which has entropy $S(\omega^{\beta})$. Thus the change in entropy of the system, $\Delta S_{\text{sys}} = S(\omega^{\beta}) - \sum_i p_i^{\beta_0}S(\omega_i^{\beta})$. 
The change in entropy of the bath is related to the heat transferred,
$\Delta S_{\text{bath}} =  \beta Q_{\text{bath}}$.
When heat transfer between the system and bath during erasure happens quasi-statically such that no work is done, one can interpret $\Delta S_{\text{bath}}$ in relation to the internal energy change of the system with the opposite sign since
\begin{equation}
\label{eq:bath_entropy}
    \Delta S_{\text{bath}} = \beta Q_{\text{bath}}= -\beta Q_{\text{sys}} = -\beta \Delta U_{\text{sys}}.
\end{equation}
where $\Delta U_{\text{sys}} = \text{Tr}[(\omega^{\beta}-\rho_\text{SS})H]$. Thus, we have $\Delta S_{\text{bath}}/\beta = E(\rho_\text{SS}) - E(\omega^{\beta}) = -\Delta U_{\text{sys}} $. We know from the previous section of the appendix that $E(\rho_\text{SS}) - E(\omega^{\beta}) = \left(S(\rho_\text{SS}) - S(\omega^{\beta}) +S(\rho_\text{SS}||\omega^{\beta})\right)/\beta$. Thus, we can express the cost of erasure as
\begin{equation}
    \Delta S(\Phi_\text{LE}) = S(\rho_\text{SS}) - \sum_i p_i^{\beta_0}S(\omega_i^{\beta}) + S(\rho_\text{SS}||\omega^{\beta}),
\end{equation}
retrieving the expression in Eq~(\ref{eq:erasure_entropy}) of the main text.

\section{Appendix C: Amplification and Mitigation}
One can arrive at various expressions to quantify the difference in internal energy by defining the free energy with respect to different temperatures. Defining free energy with respect to the initial bath temperature $T_0$ (with respect to which both $\rho_\text{SS}$ and $\omega^{\beta}$ are non-equilibrium) gives the following expressions:
\begin{subequations}
\begin{align}
E(\rho_\text{SS})  -E(\omega^{\beta_0}) &= T_0 (S(\rho_\text{SS})  - S(\omega^{\beta_0}) +S(\rho_\text{SS} ||\omega^{\beta_0}))\\
E(\omega^{\beta}) -E(\omega^{\beta_0}) &= T_0 (S(\omega^{\beta}) - S(\omega^{\beta_0}) +S(\omega^{\beta}||\omega^{\beta_0})).
\end{align}
\end{subequations}
Taking the ratio of the above two equations,
\begin{equation}
\lambda = \frac{E(\rho_\text{SS})  -E(\omega^{\beta_0})}{E(\omega^{\beta}) -E(\omega^{\beta_0})} = \frac{S(\rho_\text{SS})  - S(\omega^{\beta_0}) +S(\rho_\text{SS} ||\omega^{\beta_0})}{S(\omega^{\beta}) - S(\omega^{\beta_0}) +S(\omega^{\beta}||\omega^{\beta_0}}). \end{equation}
When $0<\beta<\beta_0$ the final energies $E(\rho_\text{SS})$ and $E(\omega^{\beta})$ are greater than the initial energy $E(\omega^{\beta_0})$. We say the s-thermalization amplifies the action of the bath when $\lambda >1$, whereas the action is mitigated when $\lambda <1$. Thus, amplification versus mitigation boundary is given by the condition $\lambda = 1$. \end{document}